\documentclass[graybox]{svmult}

\usepackage{mathptmx}       
\usepackage{helvet}         
\usepackage{courier}        
\usepackage{type1cm}        
\usepackage{makeidx}         
\usepackage{graphicx}        
\usepackage{multicol}        
\usepackage{amsmath}
\usepackage{fixltx2e}
\usepackage[bottom]{footmisc}

\makeindex             

\begin{document}

\title*{Spread of Infectious Diseases with a Latent Period}
\author{Kanako Mizuno, Kazue Kudo}
\institute{Kanako Mizuno \at Department of Computer Science, Ochanomizu
University, Tokyo, Japan \email{mizuno.kanako@is.ocha.ac.jp}
\and Kazue Kudo \at Department of Computer Science, Ochanomizu
University, Tokyo, Japan \email{kudo@is.ocha.ac.jp}}
%
%
\maketitle

\abstract{
Infectious diseases spread through human networks.
Susceptible-Infected-Removed (SIR) model is one of the epidemic models
to describe infection dynamics on a complex network connecting individuals.
In the metapopulation SIR model, each node represents a population (group)
which has many individuals. 
In this paper, we propose a modified metapopulation SIR model in which a
latent period is taken into account. We call it SIIR model.
We divide the infection period into two stages: 
an infected stage, which is the same as the previous model, 
and a seriously ill stage, in which individuals are infected 
and cannot move to the other populations.
The two infectious stages in our modified metapopulation SIR model 
produce a discontinuous final size distribution.
Individuals in the infected stage spread the disease like individuals 
in the seriously ill stage and never recover directly, which makes 
an effective recovery rate smaller than the given recovery rate.
}

%

\section{Introduction}
\label{sec:1}

Infectious diseases spread through human networks.
Susceptible-Infected-Removed (SIR) model is one of the epidemic models
to describe infection dynamics on a complex network connecting
individuals.
The ratio of the transmission rate to the recovery rate is called the basic
reproduction number $R_0$. It is the expected number of infections
caused by a typical infectious individual in a completely susceptible
population~\cite{Anderson91,Ma}.
In the standard SIR model, the outbreak occurs when $R_0>1$.
The likely magnitude of the outbreak, which is called the
expected final size of the epidemic, depends only on
$R_0$~\cite{Ma,Anderson80}.

The spread of infectious diseases also depends on human mobility.
In metapopulation SIR models, movements between different
populations (groups) are taken into
account~\cite{Keeling,Cross,Colizza}.
Each node of the metapopulation network represents a group of
individuals. Individuals can move between two nodes connected by a
link.
Although the epidemic threshold is $R_0$ in each group,
the global invasion threshold in the metapopulation system depends on
the mobility rate as well as its network structure~\cite{Cross,Colizza}.

In this paper, we propose a modified metapopulation SIR model in which a
latent period is taken into account.
Infected individuals behave like susceptible ones when they do
not feel sick. They move between linked populations and spread
diseases across different populations.
We consider that such infected individuals are in a latent period.
We assume that infected individuals become too
sick to move after the latent period.
Such ill individuals infect only the susceptible ones in the same
population.
This model is different from the SEIR model~\cite{Schwartz}, which is a
common epidemic model in which a latent period is incorporated as
an ``Exposed'' state.
However, it belongs to a family of generalized SIR models that include
multiple infectious stages~\cite{Ma}.
The two infectious stages in our modified metapopulation SIR model
produce a discontinuous final size distribution with a jump at $R_0=1$.

The rest of the paper is organized as follows.
The metapopulation SIR model and the modified SIR model are introduced
in Sec.~\ref{sec:2}. We demonstrate the discontinuous final size
distribution of the modified model in Sec.~\ref{sec:3}.
The effective recovery rate, which is different from the given recovery
rate, is estimated, and it is the key to find what causes the
discontinuity. 
Discussions and conclusions are given in Sec.~\ref{sec:4}.

\section{Model}
\label{sec:2}

First we introduce a metapopulation SIR model, which is an SIR model
that is extended to metapopulation networks.  
In the metapopulation SIR model, each node represents a population (group)
which has many individuals, and each individual is in one of three
states: $S$ (susceptible),  $I$ (infected) or $R$ (recovered). 
Individuals of state $S$ are infected by those of state $I$ in the same
population. The infection rate is given by ${\alpha}I_m/N_m$, where  
$N_m=S_m+I_m+R_m$ with $S_m$, $I_m$, and $R_m$ being the number of
susceptible, infected, and recovered individuals of population $m$,
respectively.  
In other words, the rate that $S$ becomes $I$ depends on the
transmission rate ${\alpha}$ and the
proportion of $I$ in the same population. 
The constant rate that $I$ becomes $R$, i.e., recovery rate, is defined as
${\beta}$.  
We here assume that all individuals move between the
populations connected with links in the network at a constant rate $w$.
The travel rate $w$ is the same for all the individuals.
The time evolution of the numbers of $S$, $I$ and $R$ in each population is
described by 
\begin{subequations}
\begin{eqnarray}
{\partial}_t S_n &=& - {\alpha}S_n I_n/N_n 
+ w\sum_{m}(S_m - S_n),  
\label{eq:SIR.S}\\
{\partial}_t I_n &=& {\alpha}S_n I_n / N_n - {\beta}I_n 
+ w\sum_{m}(I_m - I_n), 
\label{eq:SIR.I}\\
{\partial}_t R_n &=& {\beta}I_n + w\sum_{m}(R_m - R_n), 
\label{eq:SIR.R}
\end{eqnarray}
\label{eq:SIR}
\end{subequations}
where the summations are taken over all the populations connected with
population $n$.

Next, we divide the infection period into two stages: 
an infected stage, which is the same as the previous model, and a
seriously ill stage, in which individuals are infected and 
cannot move to the other populations.
We call this model SIIR model. In this model, each individual is in one
state of $S$ (susceptible), $H$ (infected), $I$ (seriously ill), and
$R$ (recovered). Individuals of $S$ in population $m$
are infected and become $H$ at rate 
${\alpha}(H_m+I_m)/N_m$, where $N_m=S_m+H_m+I_m+R_m$.
Individuals of $H$ become $I$ at a constant rate ${\mu}$. 
Individuals of $I$ recover and become $R$ at a rate ${\beta}$.  
In the SIIR model, individuals of $H$ move between the populations
connected with links at a rate
$w$, however, individuals of $I$ do not.  
The time evolution of the numbers of $S$, $H$, $I$ and $R$ in each
population is described by
\begin{subequations}
\begin{eqnarray}
{\partial}_t S_n &=& - {\alpha}S_n (H_n + I_n)/N_n 
+ w\sum_{m}(S_m - S_n), 
\label{eq:SIIR.S}\\
{\partial}_t H_n &=& {\alpha}S_n (H_n + I_n) / N_n - {\mu}H_n 
+ w\sum_{m}(H_m - H_n), 
\label{eq:SIIR.H}\\
{\partial}_t I_n &=& {\mu}H_n - {\beta}I_n, 
\label{eq:SIIR.I}\\
{\partial}_t R_n &=& {\beta}I_n + w\sum_{m}(R_m - R_n),
\label{eq:SIIR.R}
\end{eqnarray}
\label{eq:SIIR}
\end{subequations}
where the summations are taken over all the populations connected with
population $n$.

\section{Final Size Distribution}
\label{sec:3}

\begin{figure}[bt]
 \includegraphics[width=6.5cm]{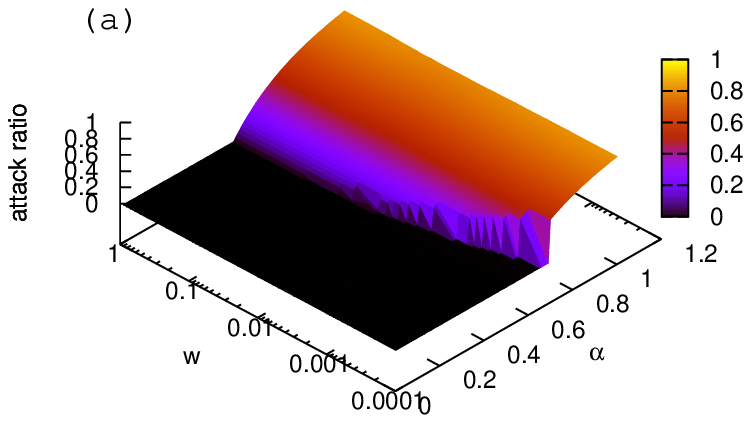}
 \includegraphics[width=6.5cm]{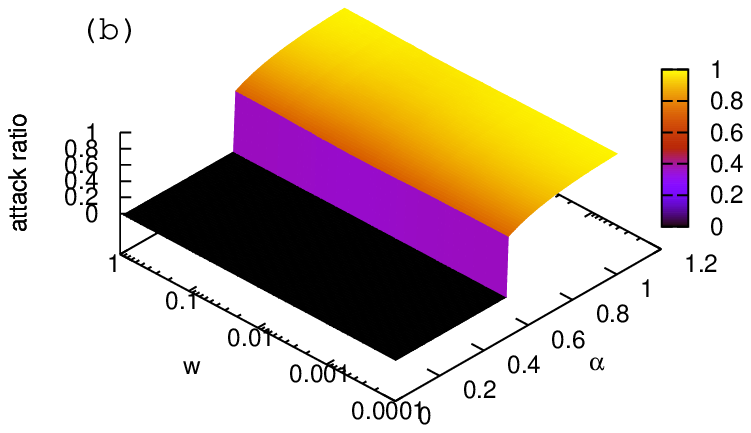}
 \caption{Final size distributions of (a) metapopulation SIR model and
 (b) metapopulation SIIR model as the function of the transmission rate
 ${\alpha}$ and the travel rate $w$.
 In both cases, the recovery rate is $\beta=0.5$. } 
\label{fig:1}       
\end{figure}

The spread of a disease is expressed by attack ratio, which is the
final proportion of $R$ when $I$ disappears in the entire
metapopulation.  
The attack ratio plotted as the function of the basic
reproduction number $\alpha/\beta$ is called a final size distribution.
The final size distributions of the SIR model and SIIR models are shown
in Fig.~\ref{fig:1}. 
In this simulation, the number of individuals in each state is taken
as a real number and the time step is discrete. We use a scale-free
network with 900 nodes, whose degree distribution is 
$P(k)\sim k^{-\gamma}$ with $\gamma=2.5$. 
The essential results do not depend on $gamma$.
In the initial state, 100 susceptible individuals belong to each node
except for one randomly selected node in which one infected
individual is included. 
The global invasion does not occur when ${\alpha}<{\beta}$ in the SIIR model
as well as the SIR model. 
The change in attack ratio is continuous at
${\alpha}={\beta}$ in the high-$w$ region in the SIR model, 
however, it is discontinuous in all region in the SIIR model. 
The shift of threshold in the low-$w$ regions of the SIR model is often
observed in 
metapopulation networks~\cite{Cross,Colizza}. 

In this paper, we focus on the discontinuous final size distribution of
the SIIR model.
The jump in the attack ratio arises from the difference between the
given recovery rate and an effective recovery rate.
In the SIIR model, individuals $H$ 
spread the disease like individuals $I$ and never become $R$ directly. 
Therefore, the effective recovery rate ${\beta}'$ is expected to be
smaller than the given recovery rate $\beta$. 

We show how to evaluate ${\beta}'$ below. 
Disregarding traveling between populations, the SIIR
model (\ref{eq:SIIR}) is rewritten as
\begin{subequations}
\begin{eqnarray}
{\partial}_t S &=& -{\alpha}S(H + I), 
\label{eq:SIIR2.S}\\
{\partial}_t H &=& {\alpha}S(H + I) - {\mu}H, 
\label{eq:SIIR2.H}\\
{\partial}_t I &=& {\mu}H - {\beta}I, 
\label{eq:SIIR2.I}\\
{\partial}_t R &=& {\beta}I,
\label{eq:SIIR2.R}
\end{eqnarray}
\label{eq:SIIR2}
\end{subequations}
where $S=S_n/N_n$, $H=H_n/N_n$, $I=I_n/N_n$ and $R=R_n/N_n$.
Combining Eqs.~(\ref{eq:SIIR2.H}) and (\ref{eq:SIIR2.I}), we have 
\begin{eqnarray*}
{\partial}_t (H+I) &=& {\alpha}S(H+I) - {\beta}'(H+I), \\
 {\beta}' &=& \frac{I}{H+I} {\beta}.
\end{eqnarray*}
We here take ${\partial}_t I=0$, which leads to $H=(\beta/\mu)I$. 
Then, the effective recovery rate is calculated as
\begin{equation}
 {\beta}' = \frac{{\mu}}{{\beta} + {\mu}} {\beta}.
\label{eq:beta_p} 
\end{equation}

\begin{figure}[tb]
\sidecaption[t]
\includegraphics[width = 7cm]{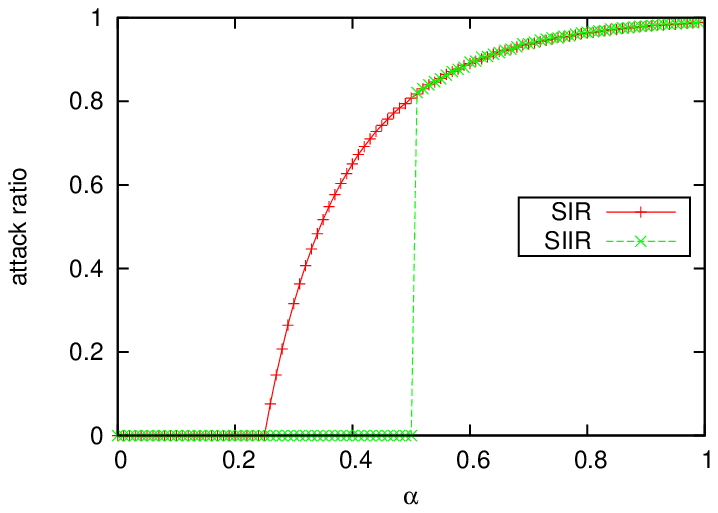}
\caption{The final size distribution as the function of the transmission
 rate $\alpha$ 
 for the SIR model with the given recovery rate ${\beta}= 0.25$ is
 compared with that for the 
 SIIR model with the effective recovery rate ${\beta}'=0.25$, 
 which is calculated from Eq.~(\ref{eq:beta_p}) with ${\beta}= 0.5$ 
 and ${\mu}= 0.5$.  
 Both curves agree in the region where $\alpha>0.5$. The travel rate
 $w=0.1$ for both 
 curves.} 
\label{fig:2}       
\end{figure}

Figure~\ref{fig:2} illustrates that the evaluation of the effective
recovery rate is appropriate.
The simulation is performed in the same network with the same initial
condition as Fig.~\ref{fig:1}. The travel rate is $w=0.1$, which is in
the high-$w$ region.  
The attack ratio for the SIIR model is calculated for ${\beta}= 0.5$
and ${\mu}= 0.5$. In this case, the effective recovery rate is
$\beta'=0.25$. 
The final size distribution for the SIR model with the given recovery rate
$\beta=0.25$ agrees with that for the SIIR model in the region where
$\alpha>0.5$.
This result implies the following.
The effective recovery rate in the SIIR model is given by $\beta'$,
however, global invasion cannot occur when $\alpha<\beta$.  
The difference between $\beta$ and $\beta'$ causes the discontinuous
final size distribution of the SIIR model. 

\begin{figure}[t]
\sidecaption[t]
\includegraphics[scale=.90]{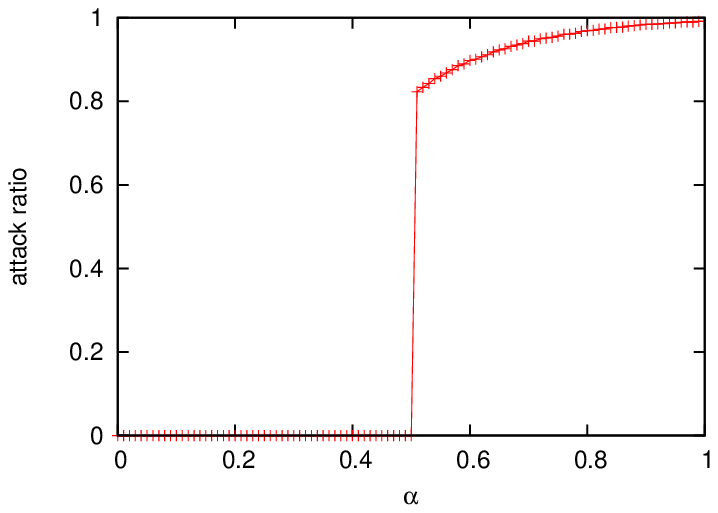}
\caption{The final size distribution of the modified SIIR model in which
 $H$ moves between populations. $\alpha$ is the transmission rate. The
 travel rate $w=0.1$.}
\label{fig:3}       
\end{figure}

Since we disregarded traveling between populations when we evaluate the
effective recovery rate, 
the assumption that $I$ is immobile should be irrelevant to the
discontinuity in the
final size distribution of the SIIR model.
We now modify the SIIR model (\ref{eq:SIIR}), replacing
Eq.~(\ref{eq:SIIR.I}) by
\begin{equation}
 {\partial}_t I_n = {\mu}H_n - {\beta}I_n + w\sum_m(I_m-I_n).
\end{equation}
Figure~\ref{fig:3} shows the final size distribution of the modified
SIIR model. 
The simulation is performed in the same conditions as Fig.~\ref{fig:2}.
The profile of the SIIR curve in Fig.~\ref{fig:2} looks the same as
the curve in Fig.~\ref{fig:3}. 
Therefore, the cause of the discontinuous final size distribution is the
division of the
infection period into two stages, and the mobility of $I$ has no effect
on the discontinuity.

\section{Discussions and Conclusions}
\label{sec:4}

The effective recovery rate $\beta'$, which is given by Eq.~(\ref{eq:beta_p}),
can be evaluated by another way.
The basic reproduction number for the generalized SIR model that
includes $n$ infectious stages is given by
\begin{equation}
 R_0=\sum_{i=1}^n\frac{\alpha_i}{\beta_i},
\label{eq:multi_R0}
\end{equation}
where $\alpha_i$ is the transmission rate of
the $i$th infectious stage, and $1/\beta_i$ is the mean duration of the
stage~\cite{Ma,Hyman}.
In our SIIR model, $\alpha_1=\alpha_2=\alpha$, $\beta_1=\mu$ and
$\beta_2=\beta$, and thus,
$R_0=\alpha/\mu+\alpha/\beta=\alpha(\mu+\beta)/(\mu\beta)$.
Therefore,
\begin{equation}
 \beta'=\frac{\alpha}{R_0}=\frac{\mu\beta}{\mu+\beta},
\end{equation}
which is the same as Eq.~(\ref{eq:beta_p}).

In conclusion, the discontinuous final size distribution in the SIIR
model is caused by the division of the infection period into two stages
and the 
fact that the global invasion cannot occur when $\alpha<\beta$. 
The final size distribution depends on the effective recovery rate
$\beta'$, and its shape coincides with that of the SIR model with a
recovery rate $\beta=\beta'$ in the region where $\alpha>\beta$. 

\begin{acknowledgement}
We would like to thank H. Takayasu and H. Nishiura for valuable
 suggestions and comments.
\end{acknowledgement}


\end{document}